\documentclass[twocolumn]{aastex631}

\usepackage{url}

\usepackage{amsmath,bm}
\usepackage{float}
\usepackage{rotating}
\usepackage{gensymb}
\usepackage{hyperref}
\usepackage{xcolor}
\usepackage{multirow}
\graphicspath{{./}{Figures/}}

\shorttitle{Evidence for Hidden Nearby Companions to Hot Jupiters}
\shortauthors{Wu et al.}

\begin{document}
\title{Evidence for Hidden Nearby Companions to Hot Jupiters}

\author[0000-0001-9424-3721]{Dong-Hong Wu}
\affiliation{Department of Physics, Anhui Normal University, Wuhu Anhui, 241000, PR, China}
\affiliation{School of Astronomy and Space Science and Key Laboratory of Modern Astronomy and Astrophysics in Ministry of Education, Nanjing University, Nanjing 210093, China}

\author[0000-0002-7670-670X]{Malena Rice}
\altaffiliation{51 Pegasi b Fellow}
\affiliation{Department of Physics and Kavli Institute for Astrophysics and Space Research, Massachusetts Institute of Technology, Cambridge, MA 02139, USA}
\affiliation{Department of Astronomy, Yale University, New Haven, CT 06511, USA}

\author[0000-0002-7846-6981]{Songhu Wang}
\affiliation{Department of Astronomy, Indiana University, Bloomington, IN 47405, USA}

\correspondingauthor{Songhu Wang}
\email{sw121@iu.edu}

\begin{abstract}
The first discovered extrasolar worlds -- giant, ``hot Jupiter'' planets on short-period orbits -- came as a surprise to solar-system-centric models of planet formation, prompting the development of new theories for planetary system evolution. The near-absence of observed nearby planetary companions to hot Jupiters has been widely quoted as evidence in support of high-eccentricity tidal migration: a framework in which hot Jupiters form further out in their natal protoplanetary disks before being thrown inward with extremely high eccentricities, stripping systems of any close-in planetary companions. In this work, we present new results from a search for transit timing variations across the full four-year \textit{Kepler} dataset, demonstrating that at least $12\pm6\%$ of hot Jupiters have a nearby planetary companion. This subset of hot Jupiters is expected to have a quiescent dynamical history such that the systems could retain their nearby companions. We also demonstrate a ubiquity of nearby planetary companions to warm Jupiters ($\geq70\pm{16}\%$), indicating that warm Jupiters typically form quiescently. We conclude by combining our results with existing observational constraints to propose an ``eccentric migration'' framework for the formation of short-period giant planets through post-disk dynamical sculpting in compact multi-planet systems. Our framework suggests that hot Jupiters constitute the natural end stage for giant planets spanning a wide range of eccentricities, with orbits that reach small enough periapses -- either from their final orbital configurations in the disk phase, or from eccentricity excitation in the post-disk phase -- to trigger efficient tidal circularization.

\end{abstract}


\section{Introduction} 
\label{section:intro}

One of the longest-standing puzzle{s} in the study of exoplanets concerns how hot Jupiters -- Jovian-sized exoplanets with orbital periods $P < 10$ days -- form and evolve (as reviewed by \citealt{Dawson2018}). Theoretical channels for hot Jupiter formation largely fall into one of the following two frameworks: 

\begin{enumerate}
    \item Dynamical high-eccentricity migration, in which cold Jupiters are launched to highly eccentric orbits through planet-planet scattering \citep{Rasio1996, Chatterjee2008}, Lidov–Kozai cycling  \citep{Wu2003, Fabrycky2007, Naoz2016_review}, or secular interactions \citep{Wu2011, Petrovich2015}, followed by tidal friction that circularizes and shrinks orbits over time. These dynamically hot processes violently deliver giant planets to their current orbits, leaving them isolated.
    \item Formation via quiescent mechanisms \citep{Goldreich_Tremaine1979,Lin_Papa1986,Lin1996,Batygin2016,Boley2016,Bailey2018}. These processes are dynamically cool, allowing nearby planetary companions to remain in the system.
\end{enumerate}

Previous population studies have suggested that hot Jupiters are rarely accompanied by nearby planetary companions analogous to those commonly observed in compact \textit{Kepler} multi-planet systems (typically with planetary radius $R_{\mathrm{p}}'\sim1-4\,{\rm R_{\oplus}}$, planetary mass $m'\sim1-10\,{\rm M_{\oplus}}$, and period ratio $P'/P\sim1.5-4$). This conclusion has been drawn based on a series of non-detections from Doppler velocimetric data \citep{Wright2009}, direct photometric searches for additional transiting planets in known hot Jupiter systems \citep{Steffen2012, Huang2016, Hord2021}, and searches for aperiodic transits of detected hot Jupiters (``transit timing variations'', or TTVs) induced by interactions with nearby companions \citep{Steffen2012, Wang2021, Ivshina2022}. 

Recently, however, a few nearby planetary companions to hot Jupiters have begun to emerge. These companions have become possible to detect through extreme-precision photometric observations from space-based observatories (e.g., WASP-47, \citealt{Becker2015}; Kepler-730, \citealt{canas2019}; TOI-1130, \citealt{Huang2020}; WASP-132, \citealt{Hord_b2022}; TOI-2000, \citealt{Sha2022}), as well as dedicated radial velocity (RV) follow-up measurements of known transiting hot Jupiters (WASP-148, \citealt{hebrard2020}). These new discoveries provide strong evidence that at least some hot Jupiters have quiescent dynamical histories. They simultaneously suggest that hot Jupiters' nearby low-mass planetary companions, which lie below the detection limits of many past surveys, may be more common than previously thought.

In this paper, we analyze TTVs across the \textit{full} 17 quarters of \textit{Kepler} data to estimate the occurrence rate of short-period gas giants with nearby planetary companions. A similar analysis of the first six quarters of \textit{Kepler} data previously found no significant TTV signals among observed hot Jupiters \citep{Steffen2012}. By contrast, our search across the complete \textit{Kepler} dataset reveals the presence of TTV signals for two hot Jupiters, as well as 14 warm Jupiters (See Figure \ref{fig:ttvsample}). 

We combine these new detections with an analysis of the underlying detection biases and completeness of the TTV method to demonstrate that at least $12\pm6\%$ of hot Jupiter systems host a nearby planetary companion. We also show that the majority of warm Jupiters host a nearby planetary companion, consistent with the results of previous work \citep{Huang2016}. Synthesizing these two results, we conclude with a unified framework for hot and warm Jupiter post-disk evolution that accounts for these elevated companion rates and that is qualitatively consistent with the full set of previous observational constraints. Our framework is rooted in both the results presented in this work and a wide range of previous constraints, which we outline in Section \ref{section:unified_framework}.

\section{A TTV Search for Kepler Jupiters}
\label{section:search_TTVs}

\subsection{Sample Selection}

We searched the complete $17$-quarter \textit{Kepler} dataset for TTV signals observed across the full sample of validated, transiting Jupiter-sized planets ($8R_{\oplus}\leq R_p\leq 16R_{\oplus}$) with impact parameter $b<0.9$. Transit parameters were drawn from \textit{Kepler} DR25 \citep{Thompson2018}, with refined stellar radii (and therefore planetary radii) from Gaia DR2 \citep{Berger2020a}. 

Our Jupiter sample (as shown in Panel $\mathbf{A}$ of Figure \ref{fig:key_result}) exhibits periods from $P=1-300\,$days, with an upper period limit set to ensure that the vast majority of Jupiters transit at least $5$ times across the $\sim$1500-day \textit{Kepler} dataset. We restricted our sample to F-, G-, and K-type main sequence stars with stellar effective temperature $4700\,{\rm K} \leq T_{\rm eff}\leq  6500\,{\rm K}$ and surface gravity ${\rm log}\,g\geq 4.0$. Our final sample includes $101$ Jupiter-sized planets.   We note that our final results depend very little on the precise sample boundaries on the Jupiters' radius, orbital period, stellar temperature, and stellar surface gravity. 

\subsection{TTV Pre-Processing}
\label{subsection:ttv_pre_processing}

We adopted transit mid-times from \citet{Holczer2016} and \citet{Rowe2015}, taking the most recently available estimate for transit mid-times included in both catalogs. We removed individual transit-mid-time measurements that were flagged as outliers in \citet{Holczer2016}, since they could lead to both false-negative and false-positive detections. Then, we re-fitted the outlier-free transit mid-times using a linear model and re-calculated the TTVs as the deviation of the observed transit mid-times from the linear orbital ephemeris. 

\subsection{TTV Detection \& Validation}
\label{TTV_detection}
We applied three previously established TTV detection techniques to our sample in search of systems with (1) excess scatter based on the modified $\chi^2$ \citep{Mazeh2013}, (2) correlated variations based on the ``alarm'' $\mathcal A$ score \citep{Tamuz:2006}, or (3) periodic signals based on the Lomb–Scargle periodogram \citep{Scargle1982, Zechmeister2009} of their TTV measurements. Each Jupiter that passed any of these three tests was identified as showing statistically significant TTVs, resulting in a set of 23 initial TTV candidates. These three tests are each described in further detail in Appendix \ref{methods:ttv_search_techniques}.

We then tested whether any of the 23 initially detected TTV candidates could have been produced by sources other than nearby planets. Potential sources of spurious TTV signals include starspots, sampling aliases, and distant planetary and stellar companions. These validation processes, which are described in Appendix \ref{method:ttv_validation}, led to the removal of six candidates from our initial TTV sample.

\begin{figure*}
    \centering
     \includegraphics[width=1.0\linewidth,trim={0cm 11cm 0 0},clip]{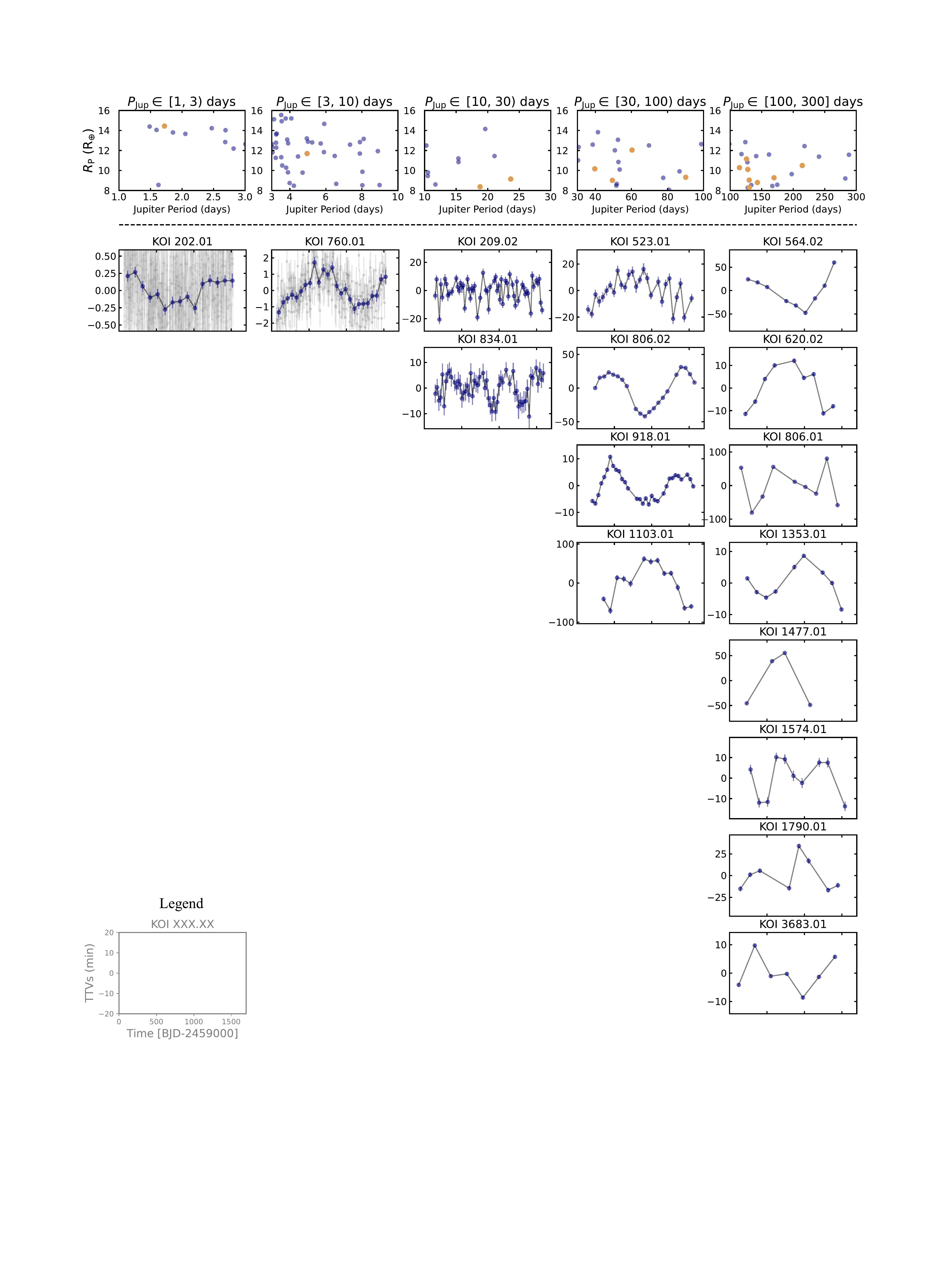}
    \caption{Detected TTV signals for the \textit{Kepler} Jupiter sample, broken into logarithmically spaced period bins. The top row shows where each system with TTVs lies in period-radius space, while the remaining panels show the observed TTVs. The legend is provided in the lower left corner for reference. The x-axis of each panel spans the same, full-time range of \textit{Kepler} primary mission observations.}
    \label{fig:ttvsample}
\end{figure*}

In total, we validated 16 out of the 23 initially identified TTV signals, demonstrating that they are most consistent with an origin from a nearby planetary companion. These systems are shown in Figure \ref{fig:ttvsample}, and the details of the 16 TTVs can be found in Table \ref{table:koi}. Two of the 16 validated TTV signals (KOI-202 and KOI-760) were observed in hot Jupiter systems, suggesting the presence of a nearby companion to the hot Jupiter in each of these systems. We examine KOI-202 and KOI-760 in greater detail in Appendix B.6 to verify the robustness of these key detections. We also encourage further studies to independently confirm the planetary origin of their TTV and/or transit signals.

Our TTV detections are in excellent agreement with previous findings that also used the full 17 quarters of \textit{Kepler} data to identify significant TTV signals \citep{Holczer2016}.  We verified that the absence of TTV detections for hot Jupiters in \citet{Steffen2012}, which incorporated only six quarters of \textit{Kepler} data, results from the differing temporal baselines used in each study  (See Appendix \ref{method:appendix_comparison} for the details of our comparison with previous studies). 

\subsection{Observed TTV Rate for \textit{Kepler} Jupiters}

\begin{figure*}
    \centering
    \includegraphics[width=1.0\linewidth]{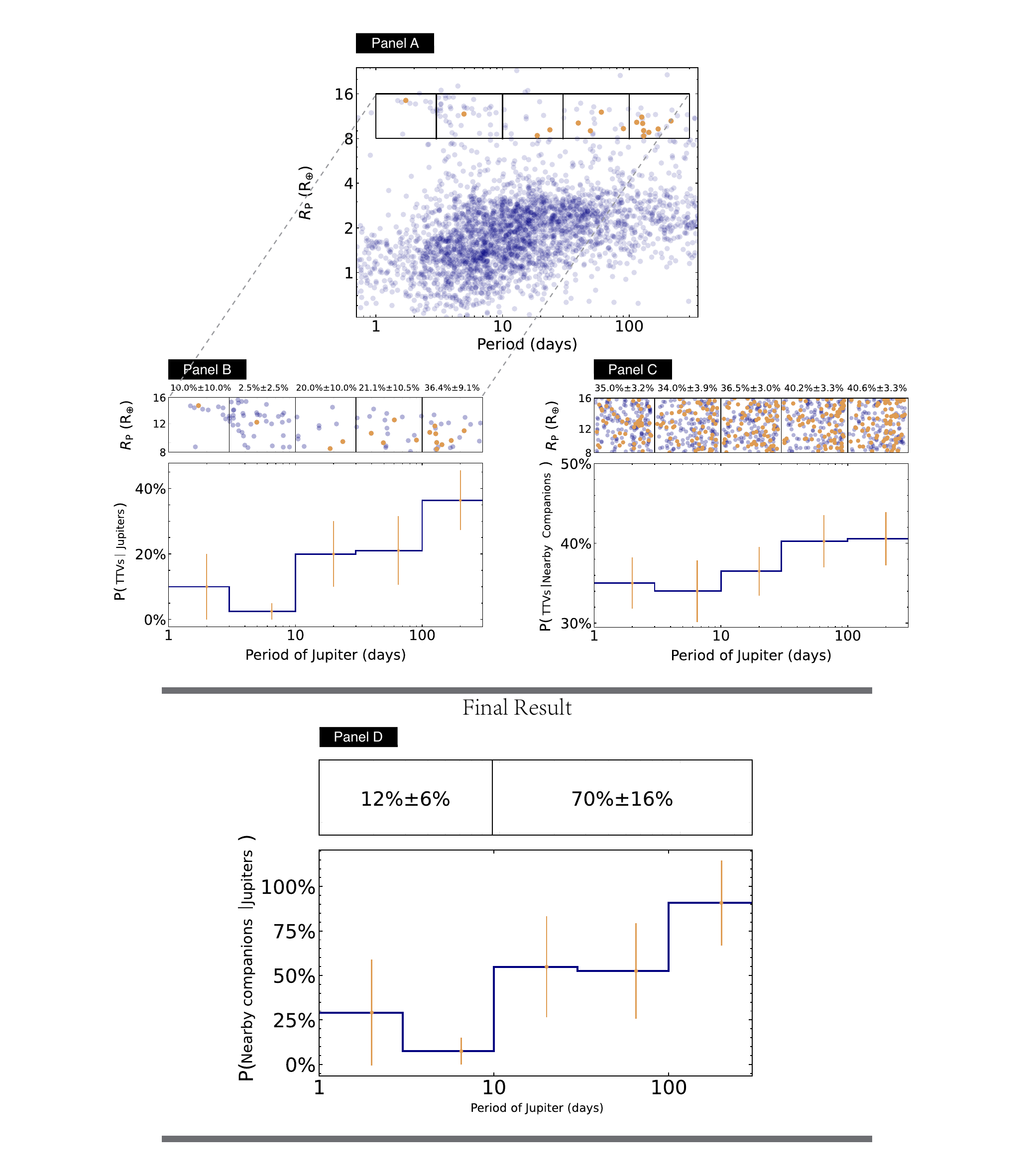}
    \caption{Summary of the observed \textit{Kepler} TTV rates, compared with simulated TTV recovery rates to provide a final occurrence rate of hot and warm Jupiters with nearby companions. \textit{Panel A:} Our full sample in period-radius space, shown in the context of \textit{Kepler} planetary candidates. \textit{Panel B:} Distribution of systems with Jupiter-sized planets that show detectable TTVs. \textit{Panel C:} Simulated recovery rate of Jupiter-massed planets with observed TTVs. \textit{Panel D:} Final combined result for the occurrence rate of Jupiter-sized planets with nearby companions as a function of the orbital period.}
    \label{fig:key_result}
\end{figure*}

To determine the fraction of \textit{Kepler} Jupiters with measured TTVs as a function of their orbital period, we divided the $101$ \textit{Kepler} Jupiter samples into five logarithmically spaced period bins between $1-300\,{\rm days}$. We then calculated the fraction P(TTVs$\mid$Jupiters) of Jupiters with detected TTVs in each bin. The resulting fractions are provided in Table~\ref{tab1} and Panel $\mathbf{B}$ of Figure \ref{fig:key_result}, where the errors are given as binomial uncertainties. The number of Jupiter-sized planets is also provided for each period bin in Table \ref{tab1}, together with the number of those planets with detected TTVs. As shown in Panel $\mathbf{B}$ of Figure \ref{fig:key_result}, we find that TTVs are significantly more common in warm Jupiter systems than in hot Jupiter systems.

\section{Completeness of the TTV Detection}

To contextualize our findings, we evaluated the detection sensitivity of the TTV technique to Jupiters' nearby planetary companions. In this section, we first describe the biases inherent to the TTV detection method through a semi-analytic approach (Section \ref{TTV_bias}). In light of these biases, we then implement a series of injection-recovery tests used to evaluate our detection efficiency as a function of the Jupiter's orbital period (Section \ref{injection_recovery test}). By accounting for the true detection sensitivity of the TTV technique, we calculate the corrective factor necessary to determine the true fraction of Jupiters with nearby planet companions.

\subsection{Semi-Analytic TTV Bias}
\label{TTV_bias}

 Given a transiting Jupiter with orbital period $P$ and mass $m$, as well as a perturbing planet with orbital period $P'$ and mass $m'$ around a host star with stellar mass $M_*$ and stellar radius $R_*$, the TTV amplitude $V_{\rm TTV}$ for the transiting Jupiter can be described as

\begin{equation}
    V_{\rm TTV} \sim P \frac{m'}{M_*} f(\frac{P'}{P},e'),\label{eq4}
\end{equation}
where $f(\frac{P'}{P}, e')$ is a function of the period ratio between the perturbing planet and transiting Jupiter ($\frac{P'}{P}$), as well as the orbital eccentricity $e'$ of the perturbing planet \citep{Holman2005, Agol2005,Lithwick2012}. This relation holds irrespective of whether the two planets are near mean-motion resonance. Equation \ref{eq4} demonstrates that, for a given period ratio $\frac{P'}{P}$ between the perturber and the Jupiter, the TTV signal is proportional to the orbital period of the transiting Jupiter ($P$). \textit{The longer the orbital period, the larger the TTV amplitude}. 

The uncertainty $\sigma_{\rm TTV}$ of a transit mid-time is given by \citet{Holman2005} as

\begin{equation}
\sigma_{\rm TTV} \sim T_{\rm 14}^{1/2}\left ( \frac{R_{\rm p}}{R_*} \right )^{-3/2} {\Gamma_*}^{-{1/2}},
\end{equation}
where $R_{\rm p}/R_*$ is the radius ratio between the transiting Jupiter and the host star, and  $\Gamma_*$ is the observed photon count rate of the target, which is proportional to its apparent brightness. $T_{14}$, the transit duration of the Jupiter, is proportional to $P^{1/3}$ such that

\begin{equation}
\sigma_{\rm TTV} \sim P^{1/6}\left (\frac{R_{\rm p}}{R_*} \right )^{-3/2}{\Gamma_*}^{-{1/2}}.\label{eq_sigttv}
\end{equation}

The signal-to-noise ratio of the TTVs, therefore, has the form

\begin{equation}
    {\rm SNR}_{\rm TTV} = \frac{V_{\rm TTV}}{\sigma_{\rm TTV}}\sim  P^{5/6}\frac{m'}{M_*}f\Big(\frac{P'}{P},e'\Big)\Big(\frac{R_{\rm p}}{R_*}\Big)^{3/2}{\Gamma_*}^{1/2},\label{eq2}
\end{equation}
which is consistent with the functional form derived in previous work \citep{Steffen2016}. 

For a given set of host star ($M_*$, $R_*$, $\Gamma_*$) and perturbing planet ($m'$, $e'$, and $\frac{P'}{P}$) properties, the SNR of observed TTVs for a transiting planet is an increasing function of its orbital period,

\begin{equation}
    \boxed{{\rm SNR}_{\rm TTV} \sim  P^{5/6}}.\label{eq1}
\end{equation}

\textit{All else equal, TTV measurements for hot Jupiters should have a lower SNR than analogous measurements of warm Jupiters}. Therefore, the low rate of detected hot-Jupiter TTVs from the \textit{Kepler} dataset may result in part from this detection bias.

\subsection{Detectability: Injection-Recovery Tests}\label{injection_recovery test}

Given the expected bias against observing TTVs in hot Jupiter systems, we next examined our detection completeness through a series of injection-recovery tests. While our analytic estimates provide a first-order approximation of the period-dependent detection bias, Equation \ref{eq1} does not account for systematic differences across our multiple detection methods. Therefore, we conducted direct simulated injection-recovery tests to more robustly characterize the completeness of our survey.

For each of the five logarithmically spaced period bins used to determine the TTV rate of \textit{Kepler} Jupiters, we generated a set of 200 trial Jupiters each with a mass of $1.0$ $M_{\rm Jup}$ around solar-mass stars ($M_*=1.0$ ${ M_{\odot}}$). The Jupiters' orbital periods were randomly drawn from a uniform distr

\begin{enumerate}
\item The mutual inclination $i_{\rm mut}$ between the two planet orbits follows a Gaussian distribution with 
{mean} $\mu_{i_{\rm mut}}=0^\circ$ and standard deviation $\sigma_{i_{\rm mut}}=1^\circ$: $i_{\rm mut} \in \mathcal{N}(\mu=0^{\circ};\sigma=1^{\circ})$.
\item Both planets follow a uniform distribution between $0^\circ$ and $360^\circ$ for the longitude of the ascending node ($\Omega,\,\Omega'$), and the initial mean anomaly ($M_{0},\,M_{0}'$); that is, $\Omega$, $\Omega'$, $M_0$, $M_0'\,  \in\,\mathcal{U}(0^{\circ};360^{\circ})$.
\item The planetary system is Hill stable ($P'/P>1.36$) \citep{gladman1993}.
\end{enumerate}

 We generated transit mid-times for each trial Jupiter by integrating the systems using the \texttt{TTVFast} code \citep{Deck2014} over a span of $1,500$ days, which is consistent with the temporal baseline of \textit{Kepler} observations. For each trial Jupiter in each period bin, the uncertainty of simulated transit mid-times ($\sigma_{\rm TTV}$) was randomly drawn from the distribution of median transit-mid-time uncertainties of the observed Jupiters in the same period bin (to account for the dependence $\sigma_{TTV} \propto P^{1/6}$; Equation \ref{eq_sigttv}).

 The simulated transit mid-times with uncertainties were then fed into our TTV detection pipeline (see Section \ref{TTV_detection}) to compute the fraction of trial Jupiters P(TTVs$\mid$Nearby Companions) showing detectable TTVs induced by the injected perturbing planets. We repeated this process $100$ times for each period bin, and using the standard deviation of the detection completeness across these $100$ trials, we found that the TTV recovery rates for each period bin are stable at the $\sim 3\%$ level (See Table~\ref{tab1} and Panel $\mathbf{C}$ of Figure \ref{fig:key_result}).

We ultimately found that the TTV detectability increases with the orbital period of the transiting Jupiters (See Table~\ref{tab1} and Panel $\mathbf{C}$ of Figure \ref{fig:key_result}), in qualitative agreement with our analytic conclusion (Equation \ref{eq1}). Although all of the TTV detection methods applied in this work are more likely to detect TTV anomalies with higher signal-to-noise ratios, they each have different biases and are sensitive to different types of signals. Therefore, while the overall TTV detectability shows an upward trend with the orbital period, it does not exhibit an exact scaling with $P^{5/6}$.

For our two shortest-period bins of hot Jupiters ($P<10$ days), only $35\%$ of the injected TTVs were identified successfully. This result reaffirms that the dearth of detected hot-Jupiter TTVs in previous studies likely results in part from detection biases. $39\%$ of the injected TTVs were recovered in warm Jupiter systems, indicating a higher but still incomplete recovery rate.

\section{Implications for Nearby Planet Occurrence}

We estimated the intrinsic fraction of Jupiters with nearby companions P(Nearby Companions$\mid$Jupiters) by dividing our newly constrained \textit{Kepler} TTV rates P(TTVs$\mid$Jupiters) by our TTV detection completeness P(TTVs$\mid$Nearby Companions). This result is provided in Table~\ref{tab1} and shown in Panel $\mathbf{D}$ of Figure \ref{fig:key_result}.
Our results demonstrate that $\geq12\pm{6}\%$ of hot Jupiters have nearby planetary companions, while the majority of warm Jupiters ($\geq70\pm{16}\%$) have nearby planetary companions. Both of these values are lower limits; that is, additional hot and/or warm Jupiters may have nearby companions that  are outside of the parameter ranges used in our injection-recovery tests and do not produce TTVs strong enough to be observable with the \textit{Kepler} dataset. 

Our work suggests a high rate of nearby companions to warm Jupiters ($\geq70\pm{16}\%$) that is consistent with, but higher than, the lower limit of $\geq58.1^{+23.9}_{-26.9}\%$ derived in \citet{Huang2016}. We also measure a relatively high rate of nearby companions to hot Jupiters ($\geq12\pm6\%$) that is consistent with the upper end of measurements reported from transit surveys within $1\sigma$: \citet{Huang2016} measure a companion rate of $0.8^{+7.7}_{-0.8}\%$, while \citet{Hord2021} measure a companion rate of $7.3^{+15.2}_{-7.3}\%$. Our work provides the first lower limit on the nearby companion rate to hot Jupiters that is nonzero within $1\sigma$.

\begin{table*}
\centering
\caption{Summary of results from our full analysis of \textit{Kepler} TTVs.}
\resizebox{\linewidth}{!}{%
\begin{tabular}{c|c|c|c|c|c|c}
\hline

Period Bins [days]& $1\leq P < 3$ & $3\leq P < 10$ & $10\leq P < 30$ & $30\leq P < 100$ & $100\leq P \leq 300.0$ &  Reference  \\\hline
\# of Jupiters&10 &40 &10 &19 &22 & \textit{Kepler} DR25, Observation\\\hline
\# of Jupiters with TTVs&1 &1 &2 &4 & 8& This work, Observation \\[5pt] \hline
P(TTVs$\mid$Jupiters) & $10.0\%\pm 10.0\%$ & $2.5\%\pm 2.5\%$ & $20.0\%\pm 10.0\%$ & $21.1\%\pm 10.5\%$ & $36.4\%\pm 9.1\%$ & This work, Observation\\[5pt] \hline
P(TTVs$\mid$Nearby Companions) & $35.0\%\pm 3.2\%$  & $34.0\%\pm 3.9\%$ & $36.5\%\pm 3.0\%$ & $40.3\%\pm 3.3\%$ & $40.6\%\pm 3.3\%$ & This work, Simulation\\[5pt] \hline

\multirow{2}{*}{\bf{P(Nearby Companions$\mid$Jupiters)}}& $\bf 29.1\%\pm 29.1\%$ & $\bf 7.4\%\pm 7.4\%$ & $\bf 54.9\%\pm 28.3\%$ & $\bf 52.5\%\pm 26.9\%$ & $\bf 90.8\%\pm 23.9\%$ &  \multirow{2}{*}{\textbf{This work, Key Result}}{} \\ 
\cline{2-6}

& \multicolumn{2}{c|}{$\bf 12\%\pm6\%$} & \multicolumn{3}{c|}{$\bf 70\%\pm16\%$} &   \\[5pt] \hline

\# of Jupiters with detected transiting companions & 0 & 1& 2 & 3 & 3 &  \textit{Kepler} DR25, Observation\\[5pt] \hline

Expected \# of Jupiters with detected transiting companions& $1\pm 1$ &$1\pm 1$ & $1\pm 1$ & $1\pm 1$ & $1\pm 1$ &  This work, Simulation\\[5pt] \hline

\end{tabular}}
\label{tab1}
\end{table*}

\section{On Mutual Inclinations}

To place our results in a broader context, we next examined whether our suggested nearby companions for Jupiters should have been detectable in the \textit{Kepler} transit survey under the assumption of coplanarity. 

We calculated the expected rate of detections by combining our predicted nearby companion rates with the well-constrained \textit{Kepler} detection efficiency for low-mass planetary companions \citep{Burke2017}. In all systems, we assumed a low mutual inclination $i_{\rm mut} \in \mathcal{N}(\mu=0^{\circ};\sigma=1^{\circ})$ between the Jupiter and its nearby companion. We adopted the same parameter distributions for companions to our Jupiters as we did in previous injection-recovery tests (See Appendix \ref{method:comparison_transit} for more details). Our results, which were calculated with the same period bins used throughout our previous analysis, are provided in the last row of Table \ref{tab1}. The expected number of \textit{Kepler} Jupiters with an additional, detectable transiting planet companion based on our TTV analysis is consistent with observations within $2\,\sigma$. 

In conclusion, our comparison between the companion rate derived from TTVs and the \textit{Kepler} transit detection indicates no compelling evidence for large mutual inclinations between \textit{Kepler} Jupiters and their nearby companions. While our sample is too small to completely rule out large mutual inclinations in these systems, it also does not indicate that they are required within the constraints of the \textit{Kepler} transit detection efficiency.

\section{Conclusions}
\label{section:conclusions}

In this paper, we have conducted a reanalysis of TTVs across the full baseline of \textit{Kepler} data, providing a new constraint on the occurrence rate of \textit{Kepler} Jupiters with nearby planetary companions. Our final results are summarized as follows:

\begin{itemize}
\item We detected TTVs for 16 Jupiter-sized planets in the \textit{Kepler} dataset -- including two hot Jupiters. We validated that these TTVs are most consistent with an origin from nearby planetary companions.

\item We demonstrated that, in addition to limitations imposed by the data's time baseline, the absence of previously observed TTV signals for \textit{Kepler} hot Jupiters \citep{Steffen2012} may, in part, result from an observational bias. This is because the TTV technique is more sensitive to warm Jupiters' nearby planetary companions than to those of hot Jupiters.

\item By correcting for TTV detection completeness, we demonstrated that a substantial fraction of hot Jupiters ($\geq12\pm{6}\%$) have nearby planetary companions, while the majority of warm Jupiters ($\geq70\pm{16}\%$) have nearby planetary companions. Here, we define ``nearby planetary companions'' as companions with radii, masses, and period ratios analogous to those within \textit{Kepler} multi-planet systems (typically with $R_{\mathrm{p}}'\sim1-4\,{\rm R_{\oplus}}$, $m'\sim1-10\,{\rm M_{\oplus}}$, and $P'/P\sim1.5-4$). Our result provides the first two-sided bound on the nearby companion rate for hot Jupiters.

\item While our sample is too small to place strong constraints on the mutual inclinations between Jupiters and their nearby planetary companions, we find no compelling evidence that large mutual inclinations are required to ensure consistency with the \textit{Kepler} sample of transit detections (agreement within 2$\sigma$).

\end{itemize}

\section{A Unified Framework for Short-Period Gas Giant Dynamical Sculpting}
\label{section:unified_framework}

In this final section, we combine our findings with other demographic properties of hot and warm Jupiters \citep{Dawson2018} to propose a framework for short-period gas giant dynamical sculpting that is qualitatively consistent with all current observational constraints. We encourage further quantitative explorations of this framework as an alternative to the standard short-period giant planet formation trichotomy of \textit{in situ} formation, disk migration, and high-eccentricity migration.

\subsection{Framework Overview}
\begin{figure*}
    \centering
    \includegraphics[width=1\linewidth, trim= 10 13 20 10, clip]{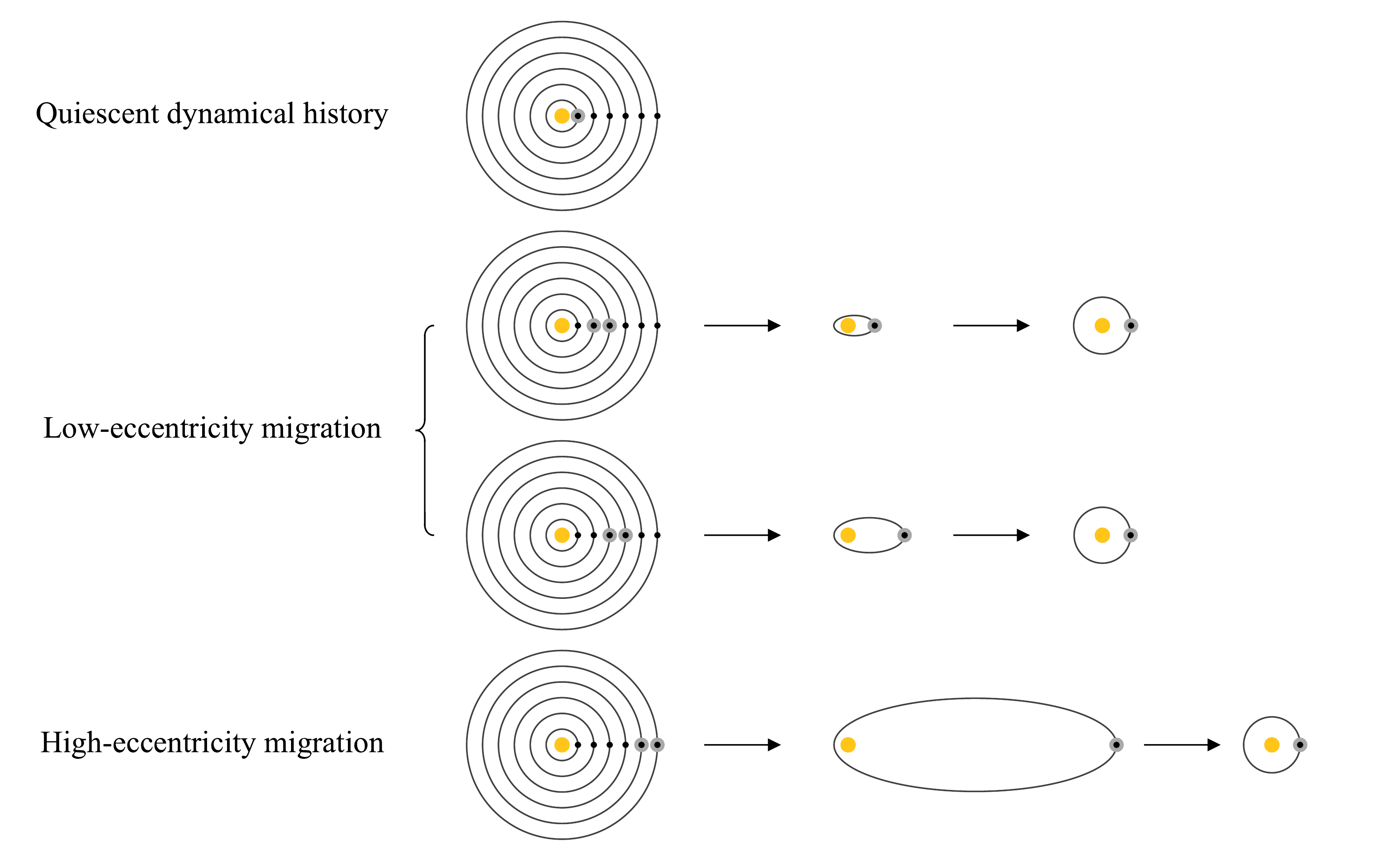}
    \caption{A selection of potential dynamical histories for hot Jupiter systems within our framework. Each path begins with a compact multi-planet system that contains at least one Jovian planet. We emphasize that the pathways shown here are intended to be representative of a broader continuum. At one extreme (top row) is a quiescent dynamical history, where hot Jupiters without strong planet-planet interactions maintain very low eccentricities ($e\sim0$), and neighboring planets remain in the system. At the other extreme (bottom row), two or more gas giants that formed further out in the disk can initialize high-eccentricity migration. Dynamical interactions between cold Jupiters launch one gas giant onto an extremely eccentric orbit, removing close-in companions from the system. The gas giant's orbit tidally circularizes over time, producing an isolated hot Jupiter. For illustrative purposes, we show two example systems that fall between these extremes, in the low-eccentricity migration regime in which some warm Jupiters may attain lower eccentricities and migrate shorter distances to ultimately become hot Jupiters. We note that some hot Jupiters that underwent eccentric migration may retain wide-orbiting companions; however, because the rate of retention for these companions is poorly constrained, we do not include them within this schematic.}
    \label{fig:HJ_formation_schematic}
\end{figure*}

In our framework, summarized in Figure \ref{fig:HJ_formation_schematic}, we propose that hot and warm Jupiters emerge as a natural outcome of post-disk dynamical sculpting of gas giants in compact multi-planet systems. The occurrence rate of these giant planets increases as a function of the amount of solid material in the disk (which is itself correlated to the mass and metallicity of the host star \citep{Fischer2005, Johnson2010}, as well as the formation location within the protoplanetary disk \citep{Wittenmyer2020, Fulton2021}). This starting point is, in part, motivated by the prevalence of ``peas-in-a-pod'' chains of super-Earths and sub-Neptunes discovered in the \textit{Kepler} dataset \citep{millholland2017kepler, weiss2018california, Goyal_Wang2022}, as well as the existence of Jovian planets observed in some compact multi-planet systems (see \citealt{Wang2017, adams2020energy}, and Figure 1 of \citealt{Dawson2019}). More importantly, however, we find that a framework based on this initialization is able to qualitatively reproduce the observed properties of hot and warm Jupiters, as is further discussed in Section \ref{subsection:congruence}.

Subsequent evolution, primarily driven by planet-planet interactions, excites the eccentricities of some giant planets. This eccentricity excitation can push the giant planets' orbits to small enough periapses that they tidally circularize and settle at shorter orbital periods. We propose that this process, which we refer to as ``eccentric migration'', occurs across a broad continuum of orbital eccentricities ranging from nearly circular ($e\sim0$) to extremely eccentric ($e>>0.9$),  with higher eccentricities required for wider-orbiting giant planets to migrate inward. For explanatory purposes, we divide the continuum of hot Jupiter eccentric migration into three regimes:

1). \textit{Hot Jupiters with quiescent histories:} Some hot Jupiters ($\geq12\pm6\%$ of the total) emerge in compact multi-planet systems without experiencing strong interactions with other planets in the same systems. As a result, these hot Jupiters maintain low eccentricities ($e\sim0$) and retain nearby planetary companions.

2). \textit{Low-eccentricity migration:} In some systems, two or more gas giants may quiescently emerge in the inner disk. After the gaseous disk has dispersed, these gas giants may experience planet-planet interactions -- including planet-planet collisions \citep{frelikh2019signatures}, scattering \citep{anderson2020situ}, or secular interactions \citep{Naoz2011, Wu2011, Petrovich2015} -- which have been shown to well reproduce the observed eccentricity distribution of close-orbiting giant planets. These interactions excite a range of eccentricities, some of which are lower than those required for conventional high-eccentricity migration but high enough to enable  the tidal circularization of shorter-period Jupiters (i.e., warm Jupiters) that ultimately become hot Jupiters. This process is responsible for producing a subset of isolated hot Jupiters.

3). \textit{High-eccentricity migration:}  In systems with two or more gas giants that emerge further out in the disk, planet-planet interactions can drive some of these gas giants to extremely eccentric orbits with periapses less than $\sim0.1\,$AU. These orbits then tidally circularize following the traditional high-eccentricity migration framework. This process is responsible for producing the remaining subset of isolated hot Jupiters.

In our framework, the $\geq12\pm6\%$ of hot Jupiters that have nearby companions emerged quiescently in compact multi-planet systems and never strongly interacted with other planets in the same systems (regime 1). The remaining $\lesssim88\%$ of hot Jupiters that lack nearby companions were delivered to their current orbits through eccentric migration (regimes 2 and 3).

We emphasize the distinction between our eccentric migration framework and that of conventional high-eccentricity migration, in which giant planets must form much farther out in the disk and therefore must reach extremely high eccentricities $e>>0.9$ to tidally circularize and become hot Jupiters. Our framework does allow for cold Jupiter migration through the high-eccentricity channel; however, it suggests that these systems are the extreme end of a continuum in which closer-in gas giants (i.e. warm Jupiters) may also attain lower eccentricities and migrate shorter distances to become hot Jupiters. Eccentric migration emerges as a natural consequence of the dynamical instabilities inherent to multiple gas giants in compact systems.

\subsection{On Hot Jupiters' Initial Formation Locations}

Because our framework emphasizes the importance of post-disk dynamical sculpting, it does not provide direct insights into the initial formation locations of extrasolar Jovian planets. In fact, subtle differences between the orbital configurations arising from large-scale disk migration and \textit{in situ} formation may be erased by this final stage of dynamical sculpting. As a result, our \textit{evolution} framework is broadly compatible with both \textit{ex situ} Jupiter formation followed by large-scale disk migration and \textit{in situ} giant planet formation. 

In the disk migration framework, our initial compact multi-planet systems with at least one gas giant may have been produced through, for example, the migration of resonant chains of planets \citep{Lee2002, bitsch2019formation}. Alternatively, \textit{in situ} Jupiter formation may occur directly in compact multi-super-Earth systems through runaway gas accretion process \citep{stevenson1982formation, Pollack1996, Boley2016, Batygin2016}. Either of these starting points may naturally produce the post-disk instability that is required in our framework to explain the observed orbital properties of hot and warm Jupiter systems.

While the initial formation locations of hot Jupiters may be obfuscated by post-disk dynamical sculpting, compositional tracers of their birthplaces may remain undisturbed. Chemical signatures of the gas giants' initial formation locations, such as the carbon-to-oxygen ratio \citep[C/O;][]{oberg2011effects} or silicon-based tracers of refractory content \citep{chachan2022breaking}, may remain detectable and can potentially be used to distinguish the initial locations of Jovian planet formation in the future (e.g. \citealt{Line2021, Reggiani2022}).

\subsection{On Formation through Stellar Kozai-Lidov}
Some hot Jupiters are located in systems with two or more known stars (e.g., HD 202772 Ab: \citealt{Wangs2019}), such that a subset of isolated hot Jupiters may have been produced through eccentric migration triggered by secular Kozai-Lidov interactions with stellar companions (e.g., HD 80606 b: \citealt{Wu2003, Naef2001}). However, hot Jupiter host stars do not have a sufficiently high rate of binary companions capable of inducing Kozai-Lidov oscillations (at most $16\pm5\%$; \citealt{ngo2015friends}) for star-planet interactions to be the dominant source of hot Jupiters. Therefore, while this formation route may contribute to the production of some isolated hot Jupiters, it cannot be the dominant channel for hot Jupiter formation.

\subsection{Congruence with Giant Planet Observations}
\label{subsection:congruence}

\subsubsection{Distant Companion Occurrence Rates}

In our framework, most hot Jupiters initially formed from compact multi-planet systems -- many of which underwent strong dynamical interactions with neighboring Jovian companions earlier in their lifetimes. Previous work has found that $70\pm8\%$ of hot Jupiters have an outer $1-13M_J$ companion at a separation between $1$ and $20$ au \citep{knutson2014friends, bryan2016statistics}. The prevalence of these distant giant companions is consistent with our framework, which suggests that planet-planet interactions may be a dominant contributor toward eccentric migration in hot Jupiter systems.

\subsubsection{Stellar Metallicity}
Past work has also demonstrated that hot Jupiters -- particularly those on eccentric orbits -- are most commonly found around metal-rich stars \citep{fischer2005planet, dawson2013giant}. A large fraction of hot Jupiters, therefore, likely originated from relatively high-metallicity protoplanetary disks that would be conducive to the growth of multiple massive planets. In our framework, the growth of multiple gas giant planets naturally triggers dynamical instabilities that lead to eccentric migration.

\subsubsection{Rarity of Super-Eccentric Jupiters}
Another set of previous results has revealed that the number of super-eccentric proto-hot Jupiters observed in the \textit{Kepler} dataset does not match the expected occurrence rate if all hot Jupiters formed through high-eccentricity migration \citep{socrates2012super, dawson2014photoeccentric, Jonathan2022}. Several potential solutions to this problem have been posed: for example, this finding may, in part, result from an observational bias in which host stars discovered by \textit{Kepler} have preferentially lower metallicities (and therefore fewer super-eccentric hot Jupiters) as compared with those detected through ground-based radial velocity surveys \citep{Guo2017}. Separate work has demonstrated that excitation of the planet's degree-2 $f$-mode can rapidly absorb orbital energy, enabling hot Jupiters with close periapse distances of $\sim$1.6 Roche radii to quickly move through the high-eccentricity stage of their migration \citep{wu2018diffusive}. Furthermore, the inflated radii of young gas giants formed through core accretion may facilitate rapid tidal migration, reducing the expected number of observable super-eccentric hot Jupiters \citep{rozner2022inflated}.

Our framework offers an alternative, relatively simple explanation for the rarity of these extremely high-eccentricity proto-hot Jupiters. The eccentric migration that we propose does not require extremely high eccentricities for all migrating Jupiters, but, rather, suggests a range of eccentricities for Jupiters migrating from different orbital distances, as well as a subset of hot Jupiters that emerged in their current orbits fully quiescently. Therefore, because not all hot Jupiters undergo high-eccentricity migration in our framework, a lower rate of super-eccentric proto-hot Jupiters would be expected.

\subsubsection{Stellar Obliquities}
Another important constraint on the formation pathways of hot Jupiters is the observed excess of spin-orbit misaligned hot Jupiters around hot stars \citep{winn2010hot, schlaufman2010evidence}. A common explanation for the difference in misalignment rates for hot and cool star systems is the distinct tidal dissipation regimes for stars above and below the Kraft break \citep[$T_{\rm eff}\sim6100$ K;][]{kraft1967studies}. That is, cool stars with convective envelopes may be capable of realigning with the hot Jupiters' orbits within the system lifetime, while hot stars with radiative envelopes are thought to be much less dissipative \citep[e.g.][]{zahn1977reprint, winn2010hot, albrecht2012obliquities, Wang_Winn2021}.

The growing sample of measured stellar obliquities in hot Jupiter systems has recently provided an opportunity to disentangle the role of eccentricity in this distribution. Of the short-period Jovian planets that are capable of tidally circularizing (with periapse distances $q < 0.1$ au), those on circular orbits ($e=0$) around cool stars tend to be aligned, whereas those that are either on eccentric orbits ($e>0.1$) or that orbit hot stars ($T_{\rm eff}>6100$ K) show an elevated rate of misalignments \citep{rice2022origins}. This suggests that obliquity excitation may be more broadly related to eccentricity excitation, as well as dynamical excitation in high-mass systems.

In our framework, the observed hot Jupiter misalignments stem from planet-planet gravitational interactions that often push Jovian planets to higher eccentricities while simultaneously misaligning the systems. Because giant planet occurrence is correlated with host star mass (and therefore stellar temperature), Jupiters around hot stars are more likely to be misaligned through planet-planet interactions than those around cool stars (as suggested in e.g. \citealt{Wang2022} \& \citealt{Yang2020}). Our framework, therefore, provides an \textit{origin} explanation for the observed difference in misalignment trends for hot and cool star systems. Planet-planet interactions could also provide an initial misalignment to trigger Kozai-Lidov interactions and produce the observed prominence of polar hot Jupiters \citep{albrecht2012obliquities, Vick2022}.

\citet{hamer2022evidence} recently showed that the hosts of misaligned hot Jupiters display a wider Galactic velocity dispersion than the hosts of aligned hot Jupiters, indicating that misaligned hot Jupiters are older than aligned hot Jupiters. Furthermore, \citet{spalding2022tidal} demonstrated that many close-orbiting hot Jupiters must have obtained their misalignments relatively late to escape realignment during the pre-main-sequence phase. These results each independently suggest that misaligned hot Jupiters typically obtain their misalignments late, after protoplanetary disk dispersal. This is an expected outcome within our framework: planet-planet interactions triggered eccentric migrations dominate the production of misaligned hot Jupiters and occur in the post-disk-dispersal phase, whereas hot Jupiters that formed quiescently would emerge in their final orbits while the disk is still present, without being misaligned in the process \citep{Sanchis-Ojeda2015, Wang2022}.

Wider-orbiting warm Jupiters with periapses too large to undergo eccentric migration and tidal realignment have so far generally exhibited spin-orbit alignment \citep{rice2022tendency}, indicating a likely quiescent formation mechanism for these systems that is consistent with their high nearby companion rate. In our framework, these warm Jupiter planets did not experience planet-planet interactions that were strong enough to trigger eccentric migration, which would have otherwise caused the removal of their nearby companions and produced large misalignments. 

\subsubsection{Orbital Eccentricities}
Our framework also addresses two critical shortcomings that arise when relying on a dichotomy of quiescent (\textit{in situ}/smooth disk migration) vs. high-eccentricity migration formation channels -- each of which cannot independently account for the observed eccentricities of short-period gas giants \citep[e.g.][]{dawson2013giant, fortney2021hot}. On the one hand, high-eccentricity migration alone cannot reproduce the large population of observed warm Jupiters with low eccentricities \citep{Petrovich_Tremaine_2016}, which, in our framework, form quiescently. On the other hand, quiescent mechanisms alone cannot account for all discovered hot Jupiters -- especially those with high eccentricities \citep{Goldreich2004}, which, in our framework, are produced by eccentric migration.

\subsection{Predictions and Implications}
Our framework offers some predictions that are testable in the immediate future. For example, hot Jupiters have been found around some pre-main-sequence T Tauri stars \citep{donati2016hot, yu2017hot}. Because of the young ages of these systems, eccentric migration likely could not deliver all hot Jupiters to their final short-period, circular orbits around these stars. Therefore, we predict a reduced occurrence rate for hot Jupiters around T Tauri stars as compared with the main sequence stars. 

Furthermore, our framework suggests that a large fraction of hot Jupiters form through planet-planet triggered eccentric migration. As a result, we expect that cold-Jupiter companions to hot Jupiters should be on more eccentric orbits than cold-Jupiter companions to warm Jupiters, most of which did not experience a violent evolutionary history.

\section{Acknowledgements}
\label{section:acknowledgements}

We are grateful to the anonymous referee for their insightful comments, which greatly improved this work. We would also like to thank Greg Laughlin, Bekki Dawson, Fred Adams, Konstantin Batygin, Eric Ford, Matthias He, Kevin Schlaufman, Bonan Pu, Jonathan Jackson, Carl Ziegler, Armaan Goyal, Jiayin Dong, and Beibei Liu for their helpful discussions. We are thankful to Xian-Yu Wang for his assistance on various aspects of this paper, and to Metrics for their contribution to polishing the figures. 

D.W. is supported by the National Natural Science Foundation of China (NSFC) (grant No. 12103003), funding from the Key Laboratory of Modern Astronomy and Astrophysics in the Ministry of Education, Nanjing University, and the Doctoral research start-up funding of Anhui Normal University. Over the duration of this project, M.R. was supported by the Heising-Simons Foundation 51 Pegasi b Fellowship and by the National Science Foundation Graduate Research Fellowship Program under Grant Number DGE-1752134. 

This research has made use of the NASA Exoplanet Archive, which is operated by the California Institute of Technology, under contract with the National Aeronautics and Space Administration under the Exoplanet Exploration Program. This work has made use of data from the European Space Agency (ESA) mission {\it Gaia} (\url{https://www.cosmos.esa.int/gaia}), processed by the {\it Gaia} Data Processing and Analysis Consortium (DPAC, \url{https://www.cosmos.esa.int/web/gaia/dpac/consortium}). Funding for the DPAC has been provided by national institutions, in particular the institutions participating in the {\it Gaia} Multilateral Agreement.

\appendix 

\section{TTV Search Techniques}
\label{methods:ttv_search_techniques}

We applied three separate methods, each described within this section, to search for significant TTV signals. Table \ref{table:koi} provides details about each TTV candidate and the detection method with which each was found.

\subsection{Detection Method 1: Excess Scatter in TTVs} 

We first identified systems with TTV scatter that was larger than expected. This ``excess scatter'' was calculated using the modified $\chi^2$ of TTVs of a system \citep{Mazeh2013,Holczer2016}, defined as:

\begin{equation}
\chi^2_{\rm modified}=N\frac{\rm (median(\{|\rm TTV_i|\}))^2}{ \rm (median(\{\sigma_{TTV_i}\}))^2},
i=1,2,\cdots, N
\end{equation}
where $N$ is the total number of TTV measurements for each target and $\sigma_{\rm TTV_i}$ is the corresponding TTV uncertainty of each measurement. Instead of taking each individual TTV measurement into account as in the conventional $\chi^2$ metric, $\chi^2_{\rm modified}$ uses the median value of the TTVs. This median value is less sensitive to outliers, such that $\chi^2_{\rm modified}$ is less sensitive to individual bad measurements. This provides a more robust detection of true TTV signals. 

 We applied a $\chi^2-$test using $\chi^2_{\rm modified}$ and obtained the corresponding $p$-value for each system, where the null hypothesis corresponds to no transit mid-time deviations from a linear transit ephemeris. We identified TTVs as significant if $p<10^{-4}$. 
\subsection{Detection Method 2: Correlated Variations in TTVs} 

The ``alarm'' $\mathcal A$ score \citep{Tamuz:2006} of a time series is sensitive to correlated variations without assuming any functional shape of the variations. It is defined as

\begin{equation}
\begin{split}
    \mathcal{A}=\frac{1}{\chi^2}\sum_{i=1}^{n}\left (\frac{\rm TTV_{i,1}}{\sigma_{\rm TTV_{i,1}}}+\cdots + \frac{\rm TTV_{i,j}}{\sigma_{\rm TTV_{i,j}}} +\cdots+\frac{\rm TTV_{i,k_i}}{\sigma_{\rm TTV_{i,k_i}}} \right )^2,
\end{split}
\end{equation}
where $\rm TTV_{i,j}$ is the $j$th TTV measurement of the $i$th ``run'' for each target, and $\sigma_{\rm TTV_{i,j}}$ is the corresponding uncertainty. A ``run'' is defined as a maximal series of consecutive TTVs with the same sign (compared with linear orbital ephemerides). The total number of runs is given by $n$, and $k_i$ is the total number of TTV measurements in the $i$th run.

 We determined the False Alarm Probability (FAP) of the calculated $\mathcal A$ score by obtaining alarm scores for 10,000 different random permutations of the same TTV series. A TTV signal was identified as significant if the FAP of the $\mathcal A$ score was less than $10^{-4}$.

\subsection{Detection Method 3: Periodic Signals in TTVs}  

The $\chi^2_{\rm modified}$ and $\mathcal A$ tests have relatively low sensitivity to periodic TTVs with small amplitudes. In order to maximize our detection yield, we also produced a Lomb-Scargle periodogram \citep{Scargle1982, Zechmeister2009} for each system. We identified the highest peak and calculated its FAP. We identified periodicities with ${\rm FAP}<10^{-4}$ as significant detections. 

\begin{table*}
\centering
\caption{KOIs with significant TTVs.}
\begin{tabular}{cccccc}
\hline
\hline
KOI & Planet Radius & Planet Period & TTV Amplitude & TTV Period & Detection Method\\\hline
 & $[R_{\rm \oplus}]$ & [days] & [min]  & [days] & \\ \hline

202.01 & 14.44 & 1.72 &0.20 &1325 & TTV periodicity \\ 
209.02 & 8.37 & 18.80 & 7.65 & 51 & $\chi^2_{\rm modified}$ \\
523.01 & 9.01 &49.4  &  10.28 & 1211 & $\chi^2_{\rm modified}$ \\ 
564.02 & 10.11 & 127.91 &  31.14& 1236 & $\chi^2_{\rm modified}$ \\ 
620.02 & 9.04 & 130.18 & 8.69& 1317 & $\chi^2_{\rm modified}$ \\ 
760.01 & 11.69 & 4.96 &  1.05 & 1051 & $\mathcal{A}$ score, TTV periodicity\\ 
806.01 & 8.80& 143.21 &  51.78&384&$\chi^2_{\rm modified}$ \\
806.02 & 12.04 & 60.32 &  23.63 & 1016 &$\chi^2_{\rm modified}$, $\mathcal{A}$ score \\ 
834.01 & 9.15 & 23.65 &  4.78 & 384 & TTV periodicity \\
918.01 & 10.17 & 39.64 &  4.75 & 835 &$\chi^2_{\rm modified}$, $\mathcal{A}$ score \\
1103.01 & 9.32 & 90.12 &  44.90 & 1068 & $\chi^2_{\rm modified}$ \\ 
1353.01 & 11.15 & 125.87 &  4.94 & 1029 & $\chi^2_{\rm modified}$\\ 
1477.01 & 9.27 & 169.50 &  47.59 & 933 & $\chi^2_{\rm modified}$ \\
1574.01 & 10.28 & 114.74 &  8.92 & 560 &$\chi^2_{\rm modified}$ \\
1790.01 & 8.31 &130.35 &  17.05 & 601 & $\chi^2_{\rm modified}$ \\
3683.01 & 10.50 & 214.31 & 5.65 & 1006 &$\chi^2_{\rm modified}$\\\hline

\end{tabular}

\label{table:koi}
\end{table*}

\section{TTV Validation}
\label{method:ttv_validation} 

\subsection{Starspots}

Spurious TTVs can be produced by the deformation of planet transits by stellar activity --- in particular, starspots \citep{2015ApJ...800..142M}. In addition to altering the measured transit mid-times, starspots also produce global modulations in each light curve as the host star rotates. This manifests in the \textit{Kepler} transit photometry as a significant local slope in the system's light curve (that is, the star's flux derivative). 

Of our 23 identified TTV systems, we found that four of them (KOI 203.01, KOI 208.01, KOI 883.01, and KOI 1458.01) show correlations between their TTVs and the local slope of their light curves. The observed correlations are likely caused by starspot crossing. Two of the four systems (KOI 203.01 and KOI 883.01) also have peak TTV periods ($11.97\pm0.9\,$days and $9.07\pm1.3\,$days, respectively) that are similar to their host stars' rotational period ($12.05$ days and $9.01$ days, respectively). As a result, we removed these four systems from our final TTV sample. The remaining 19 TTV systems do not demonstrate any periodicities related to the host stars' rotation periods or their harmonics measured from global modulations in their \textit{Kepler} light curves \citep{McQuillan_2013, 2015ApJ...800..142M, Angus2018}.

\subsection{Sampling Aliases}

Non-astrophysical effects, such as stroboscopic sampling, can also mimic significant TTVs \citep{Szab2013}. As a result, we checked whether our detected TTVs may arise as an alias of the \textit{Kepler} sampling frequency. We calculated the stroboscopic periods of the measured TTVs following the methods of previous work \citep{Szab2013}. We found that KOI 135.01 and KOI 208.01 have TTV periods ($737\pm164$ days and $373\pm25$ days, respectively) near the stroboscopic periods or their harmonics ($1532$ days and $340$ days, respectively). Therefore, we excluded these systems from our final TTV sample. 

\subsection{Stellar Companions}

We also tested the possibility that the TTV signals are produced by a stellar companion \citep{Borkovits2011}. For each system, we searched the \textit{Gaia} DR3 \citep{Gaia2016, GaiaDR3} database for all neighboring sources within $10\arcmin$, and we examined the projected separation, proper motion, and parallax of each neighboring source to search for bound companions. We adopted the same criteria for stellar companions that were outlined in Section 2 of \citet{el2021million} to develop their initial list of binary candidates, and we cross-validated our results with the \citet{el2021million} catalog.

From our search, we found no evidence that any of our observed Jupiters with significant TTVs are located in systems with a stellar companion. KOI-3683 is listed as a component of a candidate binary system in \citet{el2021million}; however, the companion candidate is located at a wide separation of 87,944 AU and is flagged with a high chance alignment probability $R=1.12$ (systems with $\sim90\%$ bound probability have $R<0.1$; see \citet{el2021million}). We conclude that this companion candidate, which would, in any case, be too wide-separation to produce the observed TTVs, is likely not bound to KOI-3683.

An additional stellar companion in a transiting planet system will induce TTVs via 1) gravitational perturbations between the stellar companion and the transiting planet \citep{Borkovits2011} and 2) the light travel time effect induced by the movement of the host star \citep{Irwin1952}. The TTV period produced by both effects should be equivalent to the orbital period of the stellar companion. None of our detected TTV signals, with typical TTV periods of several thousand days, can be produced by stellar companions unless they are nonphysically close to the planetary host stars.

\subsection{Very Distant Planetary Companions}

We then checked whether each of our identified TTV signals may instead be associated with a very distant eccentric planetary companion. Distant eccentric planetary companions can be distinguished by the non-sinusoidal TTV signals that they induce. 

 We found that two of our detected TTVs (KOI 824.01 and KOI 1474.01) are consistent with apparent non-sinusoidal variations. Previous work found that the non-sinusoidal TTVs of KOI 824.01, a $P=15$ day warm Jupiter, are likely caused by a $94\,{M_{\rm Jup}}$ companion at $a=2.8$ AU \citep{Masuda2017}, while the TTVs of KOI 1474.01, a $P=70$ day warm Jupiter, are likely caused by a $7.3\,{M_{\rm Jup}}$ companion at $a=1.7$ AU \citep{Dawson_2014}. Thus, we exclude KOI 824.01 and KOI 1474.01 to ensure that all of our validated Jupiter TTVs are consistent with an origin from a \textit{nearby} planetary companion.
 
 \subsection{Other Possibilities}
 Lastly, we searched for the signatures of other timing effects, such as apsidal precession \citep{ragozzine2009probing}, orbital decay \citep{Patra_2017}, or general relativistic effects \citep{Iorio2011}. We found that none of these possibilities are consistent with our detected TTV signals.

 \subsection{A Closer Look at KOI-202 and KOI-760}

 Since some of the key results presented within this work are largely based on the detection of TTVs for two hot Jupiter planets, we examine the robustness of these detections in greater detail here.

\subsubsection{KOI-202}
The planetary nature of KOI-202.01 (Kepler-412 b) was confirmed by \citealt{Deleuil2014} using radial velocity measurements from the SOPHIE spectrograph. KOI-202.01 is a hot Jupiter with an orbital period of $P=1.72$ days, a planetary mass of $0.939\pm{0.085} \ {\rm M_{Jup}}$, and a planetary radius of $1.325\pm0.043 \ {\rm R_{Jup}}$ \citep{Deleuil2014}. The KOI-202.01 TTV signal identified within this work was initially detected using a periodogram analysis of TTVs derived from \citet{Holczer2016} (see Section 2.3 in the main text and Section A.3 in the Appendix). To validate the robustness of the detection, we repeated the periodogram analysis using the TTVs independently derived from \citet{Rowe2015}. 

\begin{figure}[t]
    \centering
   \includegraphics[width = 0.75\textwidth, trim=0 5 1 0, clip]{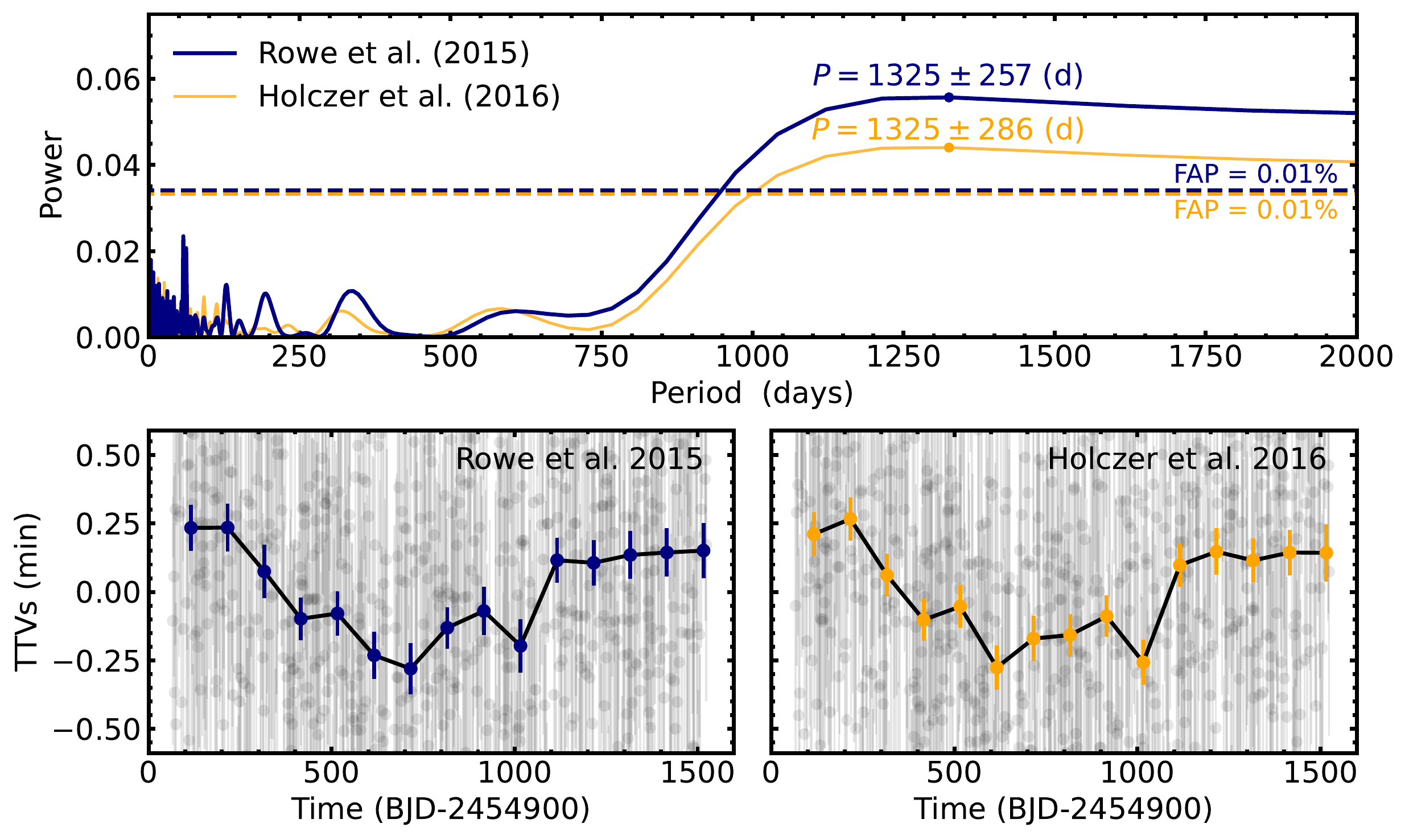}
   \caption{The upper panel displays the Generalized Lomb-Scargle Periodogram (GLS) for the TTV data of KOI 202.01 from \cite{Rowe2015} and \cite{Holczer2016}, shown in blue and yellow, respectively. The highest power and the best-fitting period are indicated by a solid dot, while the false alarm probability (FAP) level of 0.01\% is represented by a dashed line. The GLS analysis of the TTV data from these two sources suggests a strong TTV signal with a period of 1325 days. The lower panel shows the detected TTV signals for KOI 202.01, with a bin time of 100 days. The unbinned TTVs are represented by gray solid points with error bars, while the binned TTVs from \cite{Rowe2015} and \cite{Holczer2016} are depicted in blue and yellow, respectively. The binning process effectively reduces the uncertainties of the TTVs without altering the overall TTV pattern, resulting in a typical uncertainty of 0.08 minutes.\label{fig:gls}}
   
\end{figure}

The periodogram power spectra obtained from the TTVs presented in both \citet{Holczer2016} and \citet{Rowe2015} show good agreement with each other (see the top panel of Figure~\ref{fig:gls}). This agreement is not surprising given that $89.1\%$ of total transit mid-times for KOI-202, derived from the two studies, agree with each other within $0.5\,\sigma$. As shown in the bottom panel of Figure~\ref{fig:gls}, both datasets show an apparent and similar long-term TTV variation when presented with a 100-day bin (which reduces the per-point TTV uncertainty from $0.5\,$mins to $0.08\,$mins), reinforcing our confidence in the robustness of the transit mid-times to different reduction pipelines, and the robustness of the TTV signal detection.

\begin{figure}[t]
    \centering
   \includegraphics[width = 0.85\textwidth, trim=0 5 1 0, clip]{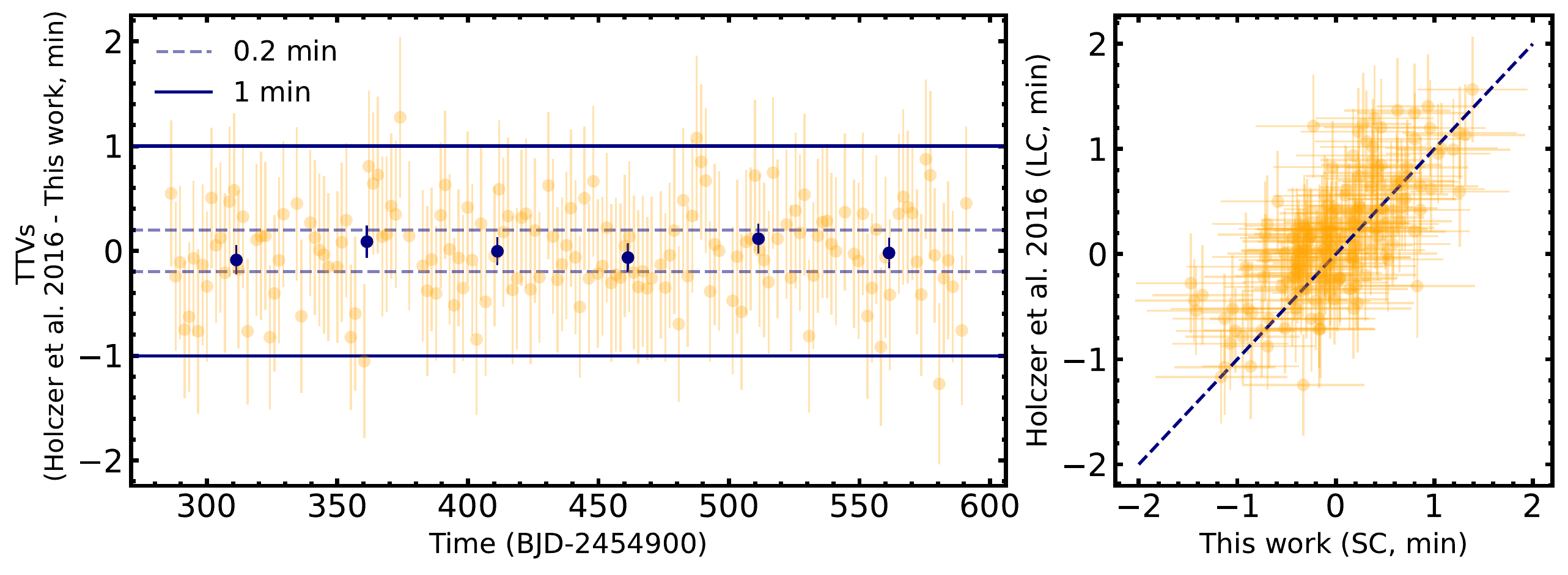}
    \caption{A comparison of TTVs derived from short-cadence data in the present study with those obtained from long-cadence data as reported in \cite{Holczer2016}. The TTV differences between \cite{Holczer2016} and this study are shown in the left panel as a function of time. The RMS of the differences was reduced from 0.44 minutes to 0.074 minutes with a 50-day bin. The right panel presents the TTVs derived from long-cadence data in \cite{Holczer2016} and the corresponding TTVs derived from short-cadence data in this study. The data points are represented by yellow dots with error bars, and the blue dashed line indicates exact agreement. \label{fig:sc_lc}}
\end{figure}

We also analyzed the short-cadence data of KOI-202 to verify whether the resulting transit mid-times are consistent with those derived from the long-cadence \textit{Kepler} data by \cite{Holczer2016}. We employed the \texttt{allesfitter} code \citep{Gunther2021} to fit each transit mid-time, independent of the \citet{Holczer2016} TTV fits. In our model, all global transit parameters were fixed to the values derived in \citealt{Deleuil2014}, and only the flux baseline and transit mid-times were allowed to vary. The uncertainties were estimated using the nested sampling method embedded in the \texttt{allesfitter} code. Since most of the \textit{Kepler} observations for KOI-202 were performed in long-cadence mode, while there were only three quarters during which data were acquired in short-cadence mode, we cannot detect the long-term variation that we have seen in the long-cadence data from short-cadence data, with a time baseline of $304.6\,$days, directly. However, as shown in Figure~\ref{fig:sc_lc}, the transit mid-times that we derived from the short-cadence data agree with those derived from long-cadence data by \cite{Holczer2016} with a root-mean-square (RMS) difference of 0.44 minutes. Furthermore, 58.5\%, 91.0\%, and 100.0\% of the long-cadence and short-cadence transit mid-times agree with each other within $0.5\,\sigma$, $1\,\sigma$, and $1.8\,\sigma$, respectively, affirming the robustness of the \citet{Holczer2016} results. The agreement between the two sets of data is reasonably good, considering the disparity in uncertainties of mid-transit times derived from the two different cadences of TTV data \citep{Carter2008}.

Finally, we modeled the TTV data from \cite{Rowe2015} and \cite{Holczer2016} using constant, quadratic, and sinusoidal functions to test the periodicity of the signal, since the TTV variation was not repeatedly observed due to its long period. The sinusoidal model was found to outperform the constant model, suggesting the existence of significant TTV variations. This is supported by the large discrepancy in the Bayesian Information Criterion  \citep[BIC,][]{schwarz1978estimating,raftery1995bayesian} between the sinusoidal and constant models, as evidenced by $\Delta \rm BIC$=40.24 for the TTVs from \citealt{Rowe2015} and $\Delta \rm BIC$=40.79 for the TTVs from \citealt{Holczer2016}. Furthermore, when comparing the sinusoidal and quadratic models, the BIC analysis also provides positive evidence favoring the sinusoidal model ($\Delta \rm BIC$=6.21 for the TTVs from \citealt{Rowe2015}, and $\Delta \rm BIC$=2.73 for the TTVs from \citealt{Holczer2016}). As a result, although the existing data time baseline is not long enough to conclusively argue that the signal must be periodic, the sinusoidal model is slightly preferred when fitting the data.

Ultimately, although the TTV amplitude of KOI-202 is small (0.2 min semi-amplitude from the best sinusoidal fitting) and the pattern has not been repeatedly observed, our analysis demonstrates that the TTV variation is robust to different reduction pipelines and is best fit by a periodic signal. We encourage follow-up photometric observations further to verify this periodicity and the planetary origin of the detected trend.

\subsubsection{KOI-760}
KOI-760.01 (KOI-760 b) is a Jupiter-sized planet candidate with an orbital period of $4.96\pm10^{-7} \,{\rm days}$. KOI-760.01 shows a significant periodic TTV signal with an amplitude of $1.05\pm0.16$ minutes and a period of $1051\pm37$ days. The detection of TTV periodicity in this system has FAP $<10^{-4}$, which cannot be reproduced by randomly shuffling the time series.

The planetary nature of KOI-760.01, however, is not confirmed, with a relatively high reported False Positive Probability of ${\rm FPP}=2.3\%$ from \citet{Morton2016} and ${\rm FPP}=7.1\%$ from \textit{Kepler} DR25 \citep{Thompson2018}. To address this point, we ran the \texttt{vespa} code \citep{Morton2012, Morton2015} on both long-cadence and short-cadence data for KOI-760 to reevaluate the likelihood that KOI-760.01 is a false positive. We found that the large FPP reported in previous work, which analyzed only long-cadence data, primarily results from the V-like shape of the phase-folded light curve. While some V-shaped transits are produced by eclipsing binaries, long-cadence data that does not sufficiently sample the light curve can also produce an apparent V-shape for a grazing planet-sized companion. However, short-cadence data should help to alleviate this ambiguity by better resolving the transit shape.

The FPP obtained from an analysis of the short-cadence data for KOI-760 is $0.8\%$, which passes the $1\%$ threshold that is used for planet validation \citep{Morton2016}.  Moreover, Robo-AO companion vetting revealed no clear stellar companions to KOI-760 \citep{Ziegler2017}. Therefore, the true FPP should be lower after accounting for the high-resolution imaging contrast curve (See \citealt{Mayo2018, canas2019}, for example). 

Like many of the planet candidates in the \textit{Kepler} Object of Interest (KOI) catalog, KOI-760.01 has not yet been directly confirmed as a planet despite undergoing a thorough vetting process. Assuming a mass of $1\,{\rm M_{\rm J}}$ for the candidate planet, the expected radial velocity semi-amplitude is approximately $120\,{\rm m\,s^{-1}}$. The planetary nature of KOI-760.01 can, therefore, be effectively verified through ground-based radial velocity follow-up observations despite the host star's relatively faint \textit{Kepler} magnitude of $K=13.8$.

\section{Comparison with Previous TTV Studies}
\label{method:appendix_comparison} 

Of the 84 Jupiter candidates included in both our analysis and in \citet{Holczer2016}, we obtained the same TTV identification on 81 of them. We identified KOI 209.02 as a significant TTV target based on its excess scatter ($\chi_{\rm modified}=163.33$, with $p<1\times 10^{-6}$ ), while it was undetected in \citet{Holczer2016}. Recent work confirmed that KOI 209.02's TTVs {are caused by an} additional transiting planet in the same system \citep{Wu_ttv_2018}. Two candidates identified as significant TTV targets in previous work, KOI 1335.01 and KOI 1552.01, did not pass any of the detection criteria that we set, nor the thresholds of the previous work; they were instead identified in previous work based on visual inspection \citep{Holczer2016}. To maintain objectivity in our results, we do not consider any candidates that were not identified by our quantitative metrics.

None of our candidates were recovered by \citet{Steffen2012}, which included only six quarters of \textit{Kepler} data and a much smaller Jupiter sample. \citet{Steffen2012} detected TTV signals for five Jupiter-sized planetary candidates; four of them (KOI 190.01, KOI 1003.01, KOI 1177.01, and KOI 1382.01), however, were later flagged as false positives in \textit{Kepler} DR25. Although KOI 137.02 (Kepler-18 c) shows a robust TTV signal, it was later confirmed to be a super-Earth \citep{Kepler_18c}.  All five of these systems were therefore excluded from our initial sample of 101 \textit{Kepler} Jupiters.

To prove that our candidates were newly recovered due to the longer temporal baseline of our dataset, we re-applied our detection pipeline to conduct the same analysis using only the first six quarters of \textit{Kepler} data. We confirmed the key result that no hot Jupiters show robust TTVs in this less extended dataset. From this analysis, we also identified four warm-Jupiter systems (KOI 209.02, KOI 806.02, KOI 918.01, and KOI 1474.01) with significant TTV signals. All four targets were unreported in the previous search because all have orbital periods longer than the upper period boundary set by that work \citep[$P=15.8$ {days};][]{Steffen2012}.

\section{Comparison with the Transit Detection Rate}
\label{method:comparison_transit} 

We examined whether these Jupiters' nearby companions should typically transit their host stars, and, correspondingly, whether these companions should have been detectable in the \textit{Kepler} dataset. We conducted an additional set of injection tests and calculated the expected recovery rates by combining our predicted nearby companion rates with the well-constrained \textit{Kepler} detection efficiency for low-mass planetary companions \citep{Burke2017}. 

We randomly assigned nearby companions to the 101 Jupiter systems in our full sample, with the companion probability set by our calculated fraction of nearby companions in each period bin (see Table 1). Injected companions were added with the following constraints:

\begin{enumerate}
    \item The mutual inclination between the companion and the Jupiter was randomly selected from a Gaussian distribution $i_{\rm mut} \in \mathcal{N}(\mu=0^{\circ};\sigma=1^{\circ})$.
    \item The radius of the planetary companion was randomly selected from the intrinsic radius distribution of \textit{Kepler} planets \citep{He2019}, most of which have $R_{\rm p}'\sim 1-4\,{\rm R_{\oplus}}$.
    
    \item The period ratio between the companion and the Jupiter was randomly selected from the intrinsic period ratio distribution of \textit{Kepler} planets given in \citet{He2019}, typically with $P'/P\sim 1.5-4$. Both inner and outer companions were considered. All planetary systems are Hill stable.
\end{enumerate}

A planet on a circular orbit is observed to transit its host star if it has an orbital inclination $90^{^\circ} - \arctan(R_{\star}/a)<i<90^{\circ} + \arctan(R_{\star}/a)$, where a perfectly edge-on orbit has $i=90^{\circ}$. The orbital inclination of the Jupiter was randomly chosen from a uniform distribution between the minimum and maximum value of this transiting range, while the orbital inclination of the companion was chosen with $i_{\rm mut} \in \mathcal{N}(\mu=0^{\circ};\sigma=1^{\circ})$ relative to the Jupiter. 

Then, we determined for each system whether the injected companion (if present) falls within the expected detection limits of \textit{Kepler}. If the planet falls outside of the inclination limits of a transiting planet, then it is not detectable. The completeness fraction for a given planet radius and orbital period is quantified as $f_{\rm tot} = f_{\rm S/N}(R,P)f_{\rm vet}(R,P)$, where $f_{\rm S/N}(R,P)$ is the \textit{Kepler} detection efficiency, while $f_{\rm vet}(R,P)$ is the vetting completeness. The fraction $f_{\rm S/N}(R, P)$ has been constrained using KeplerPORTS \citep{Burke2017}, while we calculated $f_{\rm vet}(R, P)$ on a $100\times100$ grid of logarithmically spaced radii and logarithmically spaced orbital periods following the methods of \citet{Mulders2018}. The exact value of $f_{\rm vet}(R, P)$ for each system was then interpolated from this grid.

Based on this analysis, we found that the expected number of \textit{Kepler} Jupiters with an additional, detectable transiting planet companion is consistent with observations within $2\,\sigma$ (See Table \ref{tab1}).

\bibliography{bibliography}
\bibliographystyle{aasjournal}

\end{document}